\documentclass[useAMS,usenatbib,referee]{mn2e}

\usepackage{amsmath, amssymb}
\usepackage[dvips]{graphicx}
\usepackage{epsfig}
\usepackage{color}
\usepackage{import}
\usepackage{numprint}

\newcommand{\rot}{\bmath{\nabla} \times}
\newcommand{\divg}{\bmath{\nabla}\cdot}

\newcommand{\rlight}{r_{\rm L}}
\newcommand{\Rs}{R_{\rm s}}

\newcommand{\BQ}{B_{\rm Q}}

\newcommand{\mnras}{MNRAS}

\newcommand{\apj}{ApJ}

\newcommand{\prd}{Physical Review D}

\newcommand{\ssr}{Sp. Sc. Rev.}

\title[GRQED dipole]{Quantum electrodynamical corrections to a magnetic dipole in general relativity}

\author[J. P\'etri]{J.  P\'etri$^{1}$
\thanks{E-mail: jerome.petri@astro.unistra.fr} \\
  $^{1}$Observatoire astronomique de Strasbourg, Universit\'e de Strasbourg, CNRS, UMR 7550, 11 rue de l'universit\'e, F-67000 Strasbourg, France.}

\begin{document}

\date{Accepted . Received ; in original form }

\pagerange{\pageref{firstpage}--\pageref{lastpage}} 
\pubyear{2015}

\maketitle

\label{firstpage}

\begin{abstract}
Magnetized neutron stars are privileged places where strong electromagnetic fields as high as $\BQ=4.4\times10^9$~T exist, giving rise to non-linear corrections to Maxwell equations described by quantum electrodynamics (QED). These corrections need to be included to the general relativistic (GR) description of a magnetic dipole supposed to be anchored in the neutron star. In this paper, these QED and GR perturbations to the standard flat space-time dipole are calculated to the lowest order in the fine structure constant~$\alpha_{\rm sf}$ and to any order in the ratio $\Rs/R$ where $R$ is the neutron star radius and $\Rs$ its Schwarzschild radius. Following our new 3+1~formalism developed in a previous work, we compute the multipolar non-linear corrections to this dipole and demonstrate the presence of a small dipolar~$\ell=1$ and hexapolar~$\ell=3$ component.
\end{abstract}

\begin{keywords}
  gravitation - magnetic fields - plasmas - stars: neutron - methods: analytical
\end{keywords}

\section{Introduction}

Neutron stars are exquisite objects to test our theories of gravity and electromagnetism in the strong field regime. Indeed curvature of space-time is important in the vicinity of the star due to its compactness
\begin{equation}
\label{eq:compacite}
\Xi = \frac{\Rs}{R} \approx 0.345 \, \left( \frac{M}{1.4~M_\odot} \right) \, \left( \frac{R}{12 \textrm{ km}} \right)^{-1}
\end{equation}
where $R$ is the neutron star radius $M$ its mass, $\Rs=2\,G\,M/c^2$ its Schwarzschild radius, $c$ the speed of light and $G$ the gravitational constant. Moreover, strong magnetic field are present, as high as $\BQ=4.4\times10^9$~T and even higher. These strong fields are unreachable in Earth laboratories. They act together to modify the standard expression for a pure magnetic dipole in flat space-time and weak magnetic fields $B\ll\BQ$. Whereas exact analytical solutions are known in general relativity since \cite{Ginzburg1964} for a dipole and also for multipolar terms in a spherically symmetric vacuum gravitational field as presented by \cite{1972PhRvD...6.1476W}, there is no such solutions including QED effects. The only work we are aware of is from \cite{1997JPhA...30.6475H} who already computed the corrections to a dipole to first order for any strength of the magnetic field following \cite{1936ZPhy...98..714H} effective Lagrangian. However, their description was restricted to flat space-time. The properties of strong electric and magnetic fields following the Euler-Heisenberg Lagrangian and related experiments are reviewed by \cite{2013RPPh...76a6401B}. The self-consistent Maxwell equations in curved space-time and in strong electromagnetic fields have been derived using a 3+1~formalism developed by \cite{2015MNRAS.451.3581P}. We apply these equations to the simple case of a pure static magnetic dipole. Such corrections would be most important for magnetars, i.e. neutron stars with the strongest magnetic fields known in the universe with $B\approx10-100\,\BQ$ \citep{2015RPPh...78k6901T}.

More generally, implications of strong magnetic fields to neutron star physics has been reported by \cite{2006RPPh...69.2631H}. As claimed by \cite{2003PhRvL..91g1101L}, X-ray polarization could be an efficient tool to diagnose radiation in strong electromagnetic fields. See also \cite{2000MNRAS.311..555H} and \cite{2011MNRAS.412.1381M}. Strong magnetic fields impact on atomic and molecular structure, on condensed matter, on high energy astrophysical phenomena, on accretion flows around compact objects and wave propagation as summarized by \cite{2015SSRv..191...13L}. Radiative and plasma processes in strong fields are discussed in length in \cite{meszaros1992high}.

In this paper, we solve the magnetostatic equations in general relativity with the effective Euler-Heisenberg QED Lagrangian to lowest order in spherical coordinates by using our vector spherical harmonics expansion. The magnetostatic equations and the solution techniques are reminded in Section~\ref{sec:Magnetostatic}. Results for the dipole in flat space-time and with general relativistic corrections are presented in Section~\ref{sec:Resultats}. We conclude about possible extensions of this work in the concluding remarks of Section~\ref{sec:Conclusion}.

\section{Magnetostatic equations}
\label{sec:Magnetostatic}

The general formalism to describe QED corrections including GR effects for Maxwell equations in a fixed background metric has been presented by \cite{2015MNRAS.451.3581P}. These equations simplify in the case of magnetostatic in vacuum as shown in this section. The fact that exact analytical expressions exist in GR without QED perturbations helps us greatly in looking for a series expansion in the QED Lagrangian as discussed later.

\subsection{The field equations}

To start with, we assume a non rotating dipolar magnetic field in vacuum, thus setting $\bmath E = \bmath F = \bmath J = 0$ in the equations of \cite{2015MNRAS.451.3581P}. Summarizing these simplifications in the standard notation for Maxwell equations in a medium we found the magnetostatic relations following
\begin{subequations}
\label{eq:GRQEDMaxwell}
\begin{align}
  \label{eq:Maxwell_Ampere}
  \rot \bmath{H} & = \bmath{0} \\
  \label{eq:Maxwell_Div_B}
  \divg \bmath{B} & = 0 .
 \end{align}
\end{subequations}
The constitutive relation for the magnetic field taking into account the metric and QED in the 3+1 decomposition of space-time reads
\begin{equation}
\label{eq:Constitutive}
  \mu_0 \, \mathbf H = \alpha \, \xi_1 \, \mathbf B
\end{equation}
where 
\begin{subequations}
 \begin{align}
 \alpha & = \sqrt{1 - \frac{\Rs}{r}} \\
 \xi_1 & = 1 - \frac{2\,\alpha_{\rm sf}}{45\,\upi} \, \frac{B^2}{\BQ^2} = 1 - \varepsilon \, b^2
 \end{align}
\end{subequations}
are the lapse function and the first order perturbation of the Lagrangian of the electromagnetic field for $B\ll \BQ$. See for instance \cite{2013MNRAS.433..986P, 2014MNRAS.439.1071P} for more details about the metric written in 3+1 and \cite{2015MNRAS.451.3581P} for the definition of $\xi_1$. Note also by assumption that \begin{subequations}
 \begin{align}
 \varepsilon & \equiv \frac{2\,\alpha_{\rm sf}}{45\,\upi} \ll 1 \\
 b & \equiv \frac{B}{\BQ} \ll 1.
 \end{align}
\end{subequations}
and that $\varepsilon \, b^2$ measures the strength of the perturbation accounted by QED effects. It remains a small parameter and we use $\varepsilon$ as the variable onto which we perform the series expansion.

\subsection{The field expansion}

From the general theory of vector spherical harmonics in curved space exposed in  \cite{2013MNRAS.433..986P}, in the limit of static fields, the divergencelessness constraint on the magnetic field $\bmath B$ reduces to a set of scalar functions $f^B_{\ell,m}$ such that
\begin{equation}
\label{eq:Expansion}
\mathbf{B}(r,\vartheta,\varphi,t) = \sum_{\ell=1}^\infty\sum_{m=-\ell}^\ell \rot [f^B_{\ell,m}(r,t) \, \mathbf{\Phi}_{\ell,m}]
\end{equation} 
$t$ is the coordinate time and $(r,\vartheta,\varphi)$ are the spherical coordinates.
This expansion automatically satisfies equation~(\ref{eq:Maxwell_Div_B}). Note that differential operators have to be defined with respect to the metric. The second order linear differential equations satisfied by the potentials~$f^B_{\ell,m}$ are found by inserting equation~(\ref{eq:Constitutive}) into equation~(\ref{eq:Maxwell_Ampere}).

For the special case in classical vacuum with $\varepsilon=0$, the exact solution for a static dipole in general relativity is well known since \cite{Ginzburg1964},see also \cite{2004MNRAS.352.1161R} and \cite{2013MNRAS.433..986P}, and given by
\begin{equation}
  \label{eq:DipoleSchwarzf10}
  f_{1,0}^{B({\rm dip})} = \sqrt{\frac{8\,\pi}{3}} \, B\,R^3 \, \frac{3\,r}{R_s^3} \, \left[ {\rm ln} \left( 1 - \frac{R_s}{r} \right) + \frac{R_s}{r} + \frac{R_s^2}{2\,r^2}\right] .
\end{equation}
We now seek for approximate solutions in quantum vacuum with $\varepsilon \neq 0$.

\subsection{Perturbation expansion}

We are looking for the first order corrections to this dipolar field and write therefore
\begin{equation}
 \mathbf{B} = \mathbf{B}_0 + \varepsilon \, \mathbf{B}_1
\end{equation} 
where $\mathbf{B}_0$ and $\varepsilon \, \mathbf{B}_1$ independently satisfy the divergencelessness conditions. Thus we can also expand the perturbed component according to equation~(\ref{eq:Expansion}).

The unperturbed magnetic dipole is depicted by a $0$ subscript and satisfies the relations
\begin{subequations}
\label{eq:GRQED}
\begin{align}
  \rot ( \alpha \, \bmath{B}_0 ) & = \bmath{0} \\
  \divg \bmath{B}_0 & = 0
 \end{align}
\end{subequations}
with the exact solution given by equation~(\ref{eq:DipoleSchwarzf10}). To first order in $\varepsilon$ the solution is given by
\begin{subequations}
\begin{align}
 \mu_0 \, \mathbf{H} & \approx \alpha \, ( 1 - \varepsilon \, (\mathbf{b}_0 + \varepsilon \, \mathbf{b}_1)^2 ) \, ( \mathbf{B}_0 + \varepsilon \, \mathbf{B}_1 ) \\
 & \approx \alpha \, \mathbf{B}_0 + \alpha \, \varepsilon \, ( \mathbf{B}_1 - b_0^2 \, \mathbf{B}_0) )
\end{align}
\end{subequations}
From the equilibrium condition in the unperturbed fields we get the inhomogeneous equation satisfied by the first order perturbation as
\begin{equation}
\label{eq:GREDOInhomogene}
 \rot ( \alpha \, \bmath{B}_1 ) = \rot ( \alpha \, b_0^2 \, \mathbf{B}_0)
\end{equation}
The curl on the left hand side is transformed into second order differential operators acting on $f_{\ell,m}^B$, i.e. the coefficients for the perturbation~$\bmath B_1$ such that
\begin{equation}
 \sum_{\ell=1}^\infty\sum_{m=-\ell}^\ell - \alpha \, \left[ \frac{1}{r} \, \frac{\partial}{\partial r} \left( \alpha^2\,\frac{\partial}{\partial r}(r\,f^B_{\ell,m}) \right) - \frac{\ell\,(\ell+1)}{r^2} \, f^B_{\ell,m} \right] \, \mathbf{\Phi}_{\ell,m} = \rot ( \alpha \, b_0^2 \, \mathbf{B}_0)
\end{equation}
Because the set of~$\mathbf{\Phi}_{\ell,m}$ forms an orthonormal basis functions, we project the above relation onto a particular~$\mathbf{\Phi}_{\ell,m}$ by integration over the solid angle~$d\Omega$ and get
\begin{equation}
\label{eq:EDOfBlm}
 - \alpha \, \left[ \frac{1}{r} \, \frac{\partial}{\partial r} \left( \alpha^2\,\frac{\partial}{\partial r}(r\,f^B_{\ell,m}) \right) - \frac{\ell\,(\ell+1)}{r^2} \, f^B_{\ell,m} \right] = \int \rot ( \alpha \, b_0^2 \, \mathbf{B}_0) \cdot \mathbf{\Phi}_{\ell,m} \, d\Omega
\end{equation}
It can be shown that $\rot ( \alpha \, b_0^2 \, \mathbf{B}_0)$ has only components along the vectors $\mathbf{\Phi}_{\ell,m}$. Therefore we do not miss any coefficient in the expansion of the magnetic field. Several examples are given below.

\section{Results}
\label{sec:Resultats}

Although our final goal is to compute QED corrections in general relativity, it is worthwhile to explore the Newtonian case separately. To this end, we start with exact analytical solutions to the first order QED perturbations in flat space-time and then also corrections to arbitrary order in the parameter~$\Rs/R$ in the general-relativistic case.

\subsection{Non relativistic corrections}

In flat space-time geometry, the equation satisfied by the perturbed magnetic field equation~(\ref{eq:GREDOInhomogene}) simplifies into
\begin{equation}
\label{eq:EDOInhomogene}
 \rot \bmath{B}_1 = \rot ( b_0^2 \, \mathbf{B}_0)
\end{equation}
We solve this equation with help of vector spherical harmonics and write for an aligned dipole
\begin{equation}
\rot ( b_0^2 \, \mathbf{B}_0) = \frac{B^3\,R^9}{\BQ^2 \, r^{10}} \left[ \frac{24 \, \sqrt{6\,\pi}}{5} \, \bmath{\Phi}_{1,0} + \frac{16}{5} \, \sqrt{\frac{3\,\pi}{7}} \, \bmath{\Phi}_{3,0} \right]
\end{equation}
The presence of only two terms with $(\ell,m)=\{(1,0),(3,0)\}$ is a direct consequence of the cubic dependence on the magnetic field for the source term in the right hand side in equation~(\ref{eq:EDOInhomogene}). The perturbed magnetic field~$\bmath{B}_1$ is expanded according to equation~(\ref{eq:Expansion}). Straightforward algebra detailed in \cite{2013MNRAS.433..986P} and using equation~(\ref{eq:EDOfBlm}) shows that only two coefficients are non zero. Indeed, for the aligned dipole field, these coefficients of the magnetic field expansion are given by the exact expressions
\begin{subequations}
 \begin{align}
 f^{B^{\rm NR}}_{1,0} & = - \frac{4}{15} \, \sqrt{\frac{2\,\pi}{3}} \, \frac{B^3\,R^9}{\BQ^2 \, r^8} \\
 f^{B^{\rm NR}}_{3,0} & = - \frac{4}{55} \, \sqrt{\frac{3\,\pi}{7}} \, \frac{B^3\,R^9}{\BQ^2 \, r^8}
 \end{align}
\end{subequations}
These functions correspond to a particular solution of the inhomogeneous problem. Moreover, we set the solutions of the homogeneous part to zero because they can be included in the background magnetic field before QED corrections are applied. It would simply require a rescaling of the stellar magnetic field strength. As we are not interested in this shift in magnetic field intensity, we cancel this irrelevant part. Therefore, the components of the corrections to the dipolar field are explicitly given for the $\ell=1$ mode by
\begin{subequations}
 \label{eq:Dipole10}
 \begin{align}
 B_{1,0}^{r^{\rm NR}} & = \frac{4}{15} \, \frac{B^3 \, R^9}{\BQ^2 \, r^9} \, \cos\vartheta \\
 B_{1,0}^{\vartheta^{\rm NR}} & = \frac{14}{15} \, \frac{B^3 \, R^9}{\BQ^2 \, r^9} \, \sin\vartheta
 \end{align}
\end{subequations}
and for the $\ell=3$ mode by
\begin{subequations}
 \label{eq:Hexapole30}
 \begin{align}
 B_{3,0}^{r^{\rm NR}} & = \frac{3}{110} \, \frac{B^3 \, R^9}{\BQ^2 \, r^9} \, (3\,\cos\vartheta + 5 \, \cos 3\,\vartheta) \\
 B_{3,0}^{\vartheta^{\rm NR}} & = \frac{21}{440} \, \frac{B^3 \, R^9}{\BQ^2 \, r^9} \, (\sin\vartheta + 5 \, \sin 3\,\vartheta) .
 \end{align}
\end{subequations}
The $B_{1,0}^{\varphi^{\rm NR}}$ and $B_{3,0}^{\varphi^{\rm NR}}$ components vanish everywhere. 
Our technique avoids the introduction of an artificial inner boundary as used by~\cite{1997JPhA...30.6475H}.

QED perturbations to first order give rise to a correction in the dipolar component~$(\ell,m)=(1,0)$ but also produce a hexapolar component~$(\ell,m)=(3,0)$ with a falling like $r^{-8}$ thus very different from the standard vacuum fall off as~$r^{-(\ell+1)}$. The presence of an hexapole has already been noticed by \cite{1997JPhA...30.6475H}. We also found the dependence on the cube of the magnetic moment represented by a term proportional to~²$B^3\,R^9$. Performing a judicious coordinate transform we can in principle find the components for any inclination of the dipole with respect to the $z$ axis. Nevertheless, having in mind to apply these corrections to an oblique rotating dipole, we found it more convenient to directly computed the corrections to an orthogonal rotator. Thus playing the same game for this perpendicular rotator we have
\begin{equation}
\rot ( b_0^2 \, \mathbf{B}_0) = 2 \, \frac{B^3\,R^9}{\BQ^2 \, r^{10}} \, \textrm{Re} \left[ - \frac{24\,\sqrt{3\,\pi}}{5} \, \bmath{\Phi}_{1,1} + \frac{12}{5} \, \sqrt{\frac{\pi}{7}} \, \bmath{\Phi}_{3,1} - 4 \, \sqrt{\frac{3\,\pi}{35}} \, \bmath{\Phi}_{3,3} \right]
\end{equation}
The perturbed magnetic field~$\bmath{B}_1$ is expanded according to equation~(\ref{eq:Expansion}). Again, using equation~(\ref{eq:EDOfBlm}) for the perpendicular dipole field, these coefficients are given by
\begin{subequations}
 \begin{align}
 f^{B^{\rm NR}}_{1,1} & = \frac{8\,\sqrt{3\,\pi}}{45} \, \frac{B^3\,R^9}{\BQ^2 \, r^8} \\
 f^{B^{\rm NR}}_{3,1} & = - \frac{6}{55} \, \sqrt{\frac{\pi}{7}} \, \frac{B^3\,R^9}{\BQ^2 \, r^8} \\
 f^{B^{\rm NR}}_{3,3} & = \frac{2}{11} \, \sqrt{\frac{3\,\pi}{35}} \, \frac{B^3\,R^9}{\BQ^2 \, r^8}
 \end{align}
\end{subequations}
Explicitly, the components of the corrections to the dipolar fields are for the $(\ell,m)=(1,1)$ mode
\begin{subequations}
 \begin{align}
 B_{1,1}^{r^{\rm NR}} & = \frac{4}{15} \, \frac{B^3 \, R^9}{\BQ^2 \, r^9} \, \sin\vartheta \, \cos\varphi \\
 B_{1,1}^{\vartheta^{\rm NR}} & = - \frac{14}{15} \, \frac{B^3 \, R^9}{\BQ^2 \, r^9} \, \cos\vartheta \, \cos\varphi \\
 B_{1,1}^{\varphi^{\rm NR}} & = \frac{14}{15} \, \frac{B^3 \, R^9}{\BQ^2 \, r^9} \, \sin\varphi
 \end{align}
\end{subequations}
for the $(\ell,m)=(3,1)$ mode
\begin{subequations}
 \begin{align}
 B_{3,1}^{r^{\rm NR}} & = -\frac{9}{220} \, \frac{B^3 \, R^9}{\BQ^2 \, r^9} \, (3 + 5 \, \cos2\,\vartheta) \, \sin\vartheta \, \cos\varphi \\
 B_{3,1}^{\vartheta^{\rm NR}} & = \frac{21}{\numprint{1760}} \, \frac{B^3 \, R^9}{\BQ^2 \, r^9} \, (\cos\vartheta + 15 \, \cos 3\,\vartheta) \, \cos\varphi \\
 B_{3,1}^{\varphi^{\rm NR}} & = - \frac{21}{\numprint{880}} \, \frac{B^3 \, R^9}{\BQ^2 \, r^9} \, (3 + 5 \, \cos 2\,\vartheta) \, \sin\varphi
 \end{align}
\end{subequations}
and for the $(\ell,m)=(3,3)$ mode
\begin{subequations}
 \begin{align}
 B_{3,3}^{r^{\rm NR}} & = \frac{3}{22} \, \frac{B^3 \, R^9}{\BQ^2 \, r^9} \, \sin^3\,\vartheta \, \cos 3\,\varphi \\
 B_{3,3}^{\vartheta^{\rm NR}} & = -\frac{21}{88} \, \frac{B^3 \, R^9}{\BQ^2 \, r^9} \, \cos\vartheta \, \sin^2\vartheta \, \cos 3\,\varphi \\
 B_{3,3}^{\varphi^{\rm NR}} & = \frac{21}{88} \, \frac{B^3 \, R^9}{\BQ^2 \, r^9} \, \sin^2\vartheta \, \sin 3\,\varphi
 \end{align}
\end{subequations}
These expressions can be deduced from equations~(\ref{eq:Dipole10}) and~(\ref{eq:Hexapole30}) by rotating the system of coordinates about the $y$-axis.

\subsection{General relativistic corrections}

In curved space-time geometry, the equation satisfied by the perturbed magnetic field is equation~(\ref{eq:GREDOInhomogene}). Hereto vector spherical harmonics are helpful. Getting exact analytical expression for the first order perturbation is cumbersome so we only give correction to a specified order in the ratio~$\Rs/R$. Knowing that for an aligned dipole we have
\begin{subequations}
\label{eq:RotAlphaB0}
\begin{align}
\rot ( \alpha \, b_0^2 \, \mathbf{B}_0) & = \frac{B^3\,R^9}{\BQ^2 \, r^{10}} \left[ \frac{24\,\sqrt{6\,\pi}}{5} \, \delta^R_{1,0} \, \bmath{\Phi}_{1,0} + \frac{16}{5} \, \sqrt{\frac{3\,\pi}{7}} \, \delta^R_{3,0}  \, \bmath{\Phi}_{3,0} \right] \\
\delta^R_{1,0} & = \left( 1 + \frac{79}{36} \, \frac{\Rs}{r} + \frac{\numprint{61}}{\numprint{18}} \, \frac{\Rs^2}{r^2} + \frac{\numprint{6509}}{\numprint{1440}} \, \frac{\Rs^3}{r^3} + \frac{\numprint{24959}}{\numprint{4480}} \, \frac{\Rs^4}{r^4} + \frac{\numprint{5275691}}{\numprint{806400}} \, \frac{\Rs^5}{r^5} \right) \\
\delta^R_{3,0} & = \left( 1 + \frac{3}{2} \, \frac{\Rs}{r} + \frac{\numprint{41}}{\numprint{24}} \, \frac{\Rs^2}{r^2} + \frac{\numprint{419}}{\numprint{240}} \, \frac{\Rs^3}{r^3} + \frac{\numprint{22459}}{\numprint{13440}} \, \frac{\Rs^4}{r^4} + \frac{\numprint{203621}}{\numprint{134400}} \, \frac{\Rs^5}{r^5} \right)
\end{align}
\end{subequations}
The perturbed magnetic field~$\bmath{B}_1$ is expanded according to equation~(\ref{eq:Expansion}). Straightforward algebra here again shows that only two coefficients are non zero. Nevertheless, it is impossible to get an exact analytical expression for the particular solution to equation~(\ref{eq:EDOfBlm}). We therefore resort to a series expansion in~$\Rs/R$. A particular solution to this inhomogeneous equation is found through the usual technique employing the Wronskian for the homogeneous equation as described in \cite{morse1953methodsvol1}.
Letting the new unknown function $y_{\ell,m} = r \, f^B_{\ell,m}$, we rewrite the problem as
\begin{equation}
 y_{\ell,m}''(r) + \frac{\Rs}{\alpha^2\,r^2} \, y_{\ell,m}'(r) - \frac{l\,(l+1)}{\alpha^2\,r^2} \, y_{\ell,m}(r) = - \int \frac{r}{\alpha^3} \, \rot ( \alpha \, b_0^2 \, \mathbf{B}_0) \cdot \mathbf{\Phi}_{\ell,m} \, d\Omega
\end{equation}
where the unperturbed static magnetic field is~$\bmath B_0$ and the right-hand side is known explicitly according to equation~(\ref{eq:RotAlphaB0}).

The particular solution can be represented by an integral including the Wronskian and two linearly independent solutions of the homogeneous equation. For the aligned dipole field, the coefficients~$f^B_{\ell,m}$ of the magnetic field expansion are given by
\begin{subequations}
 \begin{align}
 f^B_{1,0} & = f^{B^{\rm NR}}_{1,0} \, \delta^{\rm L}_{1,0} \\
 f^B_{3,0} & = f^{B^{\rm NR}}_{3,0} \, \delta^{\rm L}_{3,0} \\ 
 \delta^{\rm L}_{1,0} & = 1 + \frac{417}{140} \, \frac{\Rs}{r} + \frac{\numprint{7011}}{\numprint{1232}} \, \frac{\Rs^2}{r^2} + \frac{\numprint{19917}}{\numprint{2240}} \, \frac{\Rs^3}{r^3} + \frac{\numprint{225441}}{\numprint{18200}} \, \frac{\Rs^4}{r^4} + \frac{\numprint{691443}}{\numprint{43120}} \, \frac{\Rs^5}{r^5}  \\
 \delta^{\rm L}_{3,0} & = 1 + \frac{\numprint{151}}{\numprint{60}} \, \frac{\Rs}{r} + \frac{\numprint{163}}{\numprint{39}} \, \frac{\Rs^2}{r^2} + \frac{\numprint{73667}}{\numprint{12740}} \, \frac{\Rs^3}{r^3} + \frac{\numprint{3680413}}{\numprint{509600}} \, \frac{\Rs^4}{r^4} + \frac{\numprint{9544139}}{\numprint{1128960}} \, \frac{\Rs^5}{r^5} 
 \end{align}
\end{subequations}
The superscript $^{\rm L}$ stands for the longitudinal part, the radial component.
Explicitly, the components of the corrections to the dipolar fields are for the $\ell=1$ mode
\begin{subequations}
 \begin{align}
 B_{1,0}^r & = B_{1,0}^{r^{\rm NR}} \, \delta^{\rm L}_{1,0} \\
 B_{1,0}^\vartheta & = B_{1,0}^{\vartheta^{\rm NR}} \, \delta^{\rm T}_{1,0} \\
 \delta^{\rm T}_{1,0} & = 1 + \frac{\numprint{1423}}{\numprint{490}} \, \frac{\Rs}{r} + \frac{\numprint{236713}}{\numprint{43120}} \, \frac{\Rs^2}{r^2} + \frac{\numprint{368927}}{\numprint{43120}} \, \frac{\Rs^3}{r^3} + \frac{\numprint{597977}}{\numprint{50050}} \, \frac{\Rs^4}{r^4} + \frac{\numprint{4881733783}}{\numprint{313913600}} \, \frac{\Rs^5}{r^5}
 \end{align}
\end{subequations}
and for the $\ell=3$ mode they are
\begin{subequations}
 \begin{align}
 B_{3,0}^r & = B_{3,0}^{r^{\rm NR}} \, \delta^{\rm L}_{3,0} \\
 B_{3,0}^\vartheta & = B_{3,0}^{\vartheta^{\rm NR}} \, \delta^{\rm T}_{3,0} \\
 \delta^{\rm T}_{3,0} & = 1 + \frac{\numprint{499}}{\numprint{210}} \, \frac{\Rs}{r}  + \frac{\numprint{41611}}{\numprint{10920}} \, \frac{\Rs^2}{r^2} + \frac{\numprint{5513087}}{\numprint{1070160}} \, \frac{\Rs^3}{r^3} + \frac{\numprint{270893431}}{\numprint{42806400}} \, \frac{\Rs^4}{r^4} + \frac{\numprint{625810279}}{\numprint{85612800}} \, \frac{\Rs^5}{r^5}
\end{align}
\end{subequations}
The superscript $^{\rm T}$ stands for the transverse part, the spherical components $(\vartheta, \varphi)$.

We can play the same game for the orthogonal rotator. Knowing that for an orthogonal dipole we have
\begin{subequations}
\begin{align}
\rot ( \alpha \, b_0^2 \, \mathbf{B}_0) & = \frac{B^3\,R^9}{\BQ^2 \, r^{10}}  \left[ - \frac{48\,\sqrt{3\,\pi}}{5} \, \delta^R_{1,0} \, \bmath{\Phi}_{1,1} + \frac{24}{5} \, \sqrt{\frac{\pi}{7}} \, \delta^R_{3,0} \, \bmath{\Phi}_{3,1} - 8 \, \sqrt{\frac{3\,\pi}{35}} \, \delta^R_{3,0} \, \bmath{\Phi}_{3,3} \right]
\end{align}
\end{subequations}
The perturbed magnetic field~$\bmath{B}_1$ is expanded according to equation~(\ref{eq:Expansion}). Straightforward algebra detailed in \cite{2013MNRAS.433..986P} shows that only three coefficients are non zero. Indeed, for the perpendicular dipole field, these coefficients are given by
\begin{subequations}
 \begin{align}
 f^B_{1,1} & = f^{B^{\rm NR}}_{1,1} \, \delta^{\rm L}_{1,0} \\
 f^B_{3,1} & = f^{B^{\rm NR}}_{3,1} \, \delta^{\rm L}_{3,0} \\
 f^B_{3,3} & = f^{B^{\rm NR}}_{3,3} \, \delta^{\rm L}_{3,0}
 \end{align}
\end{subequations}
Explicitly, the components of the corrections to the dipolar fields are for the $(\ell,m)=(1,1)$ mode
\begin{subequations}
 \begin{align}
 B_{1,1}^r & = B_{1,1}^{r^{\rm NR}} \, \delta^{\rm L}_{1,0} \\
 B_{1,1}^\vartheta & = B_{1,1}^{\vartheta^{\rm NR}} \, \delta^{\rm T}_{1,0} \\
 B_{1,1}^\varphi & = B_{1,1}^{\varphi^{\rm NR}} \, \delta^{\rm T}_{1,0}
 \end{align}
\end{subequations}
for the $(\ell,m)=(3,1)$ mode
\begin{subequations}
 \begin{align}
 B_{3,1}^r & = B_{3,1}^{r^{\rm NR}} \, \delta^{\rm L}_{3,0} \\
 B_{3,1}^\vartheta & = B_{3,1}^{\vartheta^{\rm NR}} \, \delta^{\rm T}_{3,0} \\
 B_{3,1}^\varphi & = B_{3,1}^{\varphi^{\rm NR}} \, \delta^{\rm T}_{3,0}
\end{align}
\end{subequations}
and for the $(\ell,m)=(3,3)$ mode
\begin{subequations}
 \begin{align}
 B_{3,3}^r & = B_{3,3}^{r^{\rm NR}} \, \delta^{\rm L}_{3,0} \\
 B_{3,3}^\vartheta & = B_{3,3}^{\vartheta^{\rm NR}} \, \delta^{\rm T}_{3,0} \\
 B_{3,3}^\varphi & = B_{3,3}^{\varphi^{\rm NR}} \, \delta^{\rm T}_{3,0} .
\end{align}
\end{subequations}
The $\ell=3$ modes possess the same GR corrections as expected from the geometric transform through a rotation about the $y$-axis. The orthogonal rotator does not lead to new radial dependences of the magnetic field components compared to the aligned rotator. This is expected because through a judicious coordinate system transform we can retrieve the latter geometry. In realistic neutron stars, the ratio $\Rs/R$ is at most~$0.5$ such that the series of terms $S_n = (\Rs/R)^n$ converge to zero geometrically. As the series in $\delta$ only converges slowly, we had to take several terms in the summation to reach an acceptable accuracy. The rate of convergence depends strongly on the stellar compactness $\Rs/R \lesssim 0.5$.

\section{Discussion}

The above results show the effect of vacuum polarization on the structure of a purely magnetic field. In a real astrophysical context such as magnetospheres of magnetars, the strength of the magnetic field at the surface can largely exceed the quantum critical field~$\BQ$. The associated electromagnetic stress-energy tensor acts as a supplementary source for the gravitational field. However even for such extreme field strengths the generation of gravitational fields remains marginal. Indeed the induced curvature in the metric geometry can be estimated by the ratio of magnetic energy density over rest mass energy density of the star by the expression
\begin{equation}
 \frac{2\,\upi}{3\,\mu_0} \, \frac{B^2\,R^3}{M\,c^2} \approx 3.1 \times 10^{-10} \, \left( \frac{B}{\BQ} \right)^2 \, \left( \frac{R}{12 \textrm{ km}} \right)^{3} \, \left( \frac{M}{1.4~M_\odot} \right)^{-1}
\end{equation}
assuming constant and homogeneous magnetization and density inside the star. Thus space-time distortions from the electromagnetic field remain rather weak compared to equation~(\ref{eq:compacite}).

Moreover, rotation of the magnet, assuming to be a perfect conductor inside the star, induces an electric field at the surface of the order
\begin{equation}
 \frac{E}{c\,B} = \frac{R}{\rlight} = \frac{2\,\upi\,R}{c\,P} \approx 2.5 \times 10^{-4} \, \left( \frac{R}{12 \textrm{ km}} \right) \, \left( \frac{P}{1~s} \right)^{-1}
\end{equation}
For magnetars, the observed period~$P$ of rotation is more than one second (between 2-12~seconds according to \cite{2015RPPh...78k6901T}). This small rotation rate implies a large light-cylinder radius of more than $9\times10^7$~meters and a ratio $R/\rlight \lesssim 1.3 \times 10^{-4}$. Therefore we do not expect the electric field to drastically change the properties of quantum vacuum. Moreover, in the slow rotation approximation, corrections to the magnetic field inside the light-cylinder are of the order~$r/\rlight$, see for instance appendix~B of \cite{2015MNRAS.450..714P}. Consequently, rotation of the neutron star will not affect the multipole geometry described in the previous section. In any way, the field of a magnetar is significantly affected by currents in its magnetosphere. In the standard picture pulsar magnetospheres, the current induced by the corotation of charges distorts the electromagnetic field to the same order of magnitude than rotation in vacuum alone. However, in the twisted magnetosphere of magnetars, the supporting currents for the magnetic twists are much larger than the currents induced by the corotation charge density \citep{2007ApJ...657..967B}. Such high current flows are evidenced by their non-thermal emission properties and should be taken into account for an accurate description of the magnetic topology of neutron star magnetic fields.

\section{Conclusion}
\label{sec:Conclusion}

Strong electromagnetic fields in neutron stars and especially in magnetars are at the heart of pulsar machinery, the crucial electron-positron pair creation scenario and the related radiative processes. If the emission observed at Earth emanates from the base of the magnetosphere, close to the stellar surface, we should have indirect evidence of these strong fields. The magnetic topology is also of great importance to understand the observations. In this paper we have shown that QED corrections induce multipolar components in the magnetic field. We gave exact analytical expressions for flat space-time and a power expansion series in $\Rs/R$ in curved space-time. The strength of the corrections decrease very steeply with radius at least as the ninth power of~$r$, $B_{\rm QED} \propto r^{-9}$. Therefore QED corrections to the field can only lead to palpable effects in the vicinity of the neutron star surface. 

What would happen if this GRQED dipole is put into rotation and emits a large amplitude electromagnetic wave in vacuum? Would the classical magnetodipole formula still be valid or should we expect an enhancement in the radiating fields. We answer quantitatively this question in another paper by performing time-dependent numerical simulations of a inclined rotating dipole in curved space-time including vacuum polarization. We also plane to extend these corrections to other magnetic field topologies such as a quadrupole, a hexapole or an octupole as well as to an arbitrary strong static magnetic field with any value of $B$, not necessarily $B\ll\BQ$, using for instance an analytical formula for the effective Lagrangian derived by \cite{1997PhRvD..55.2449H}.

\section*{Acknowledgements}

I am grateful to the referee for helpful comments. This work has been supported by the French National Research Agency (ANR) through the grant No. ANR-13-JS05-0003-01 (project EMPERE).


\label{lastpage}

\end{document}